\documentclass[aps,twocolumn,showpacs,amsmath,amssymb,superscriptaddress,final,prb,a4paper,10pt,longbibliography]{revtex4-1}
\usepackage[utf8]{inputenc} %[latin1]
\usepackage{dcolumn}  % For aligning table column on decimal point
\usepackage{nicefrac}   % For nice small fractions
\usepackage[pdftex]{color}
\usepackage{bm}
\usepackage[charter,greekuppercase=italicized]{mathdesign}
\usepackage{charter}

\usepackage{soul} % For easier highlights use \hl{highlighted words}

\usepackage[colorlinks,plainpages=false,linkcolor=blue,urlcolor=blue,citecolor=blue,pdfpagemode=UseNone,pdfstartview=FitBH]{hyperref}

\definecolor{red}{rgb}{0.85,.1,0}
\definecolor{orange}{rgb}{1,0.5,0}
\usepackage{graphicx}  % For including figure files
\graphicspath{{./}{figure/}}
\DeclareGraphicsExtensions{.eps,.png,.pdf,.jpg}

\newcommand{\sfmaof}{Sm\-Fe$_{1-x}$Mn$_{x}$\-As\-O$_{0.88}$\-F$_{0.12}$}

\newcommand{\lfmao}{La\-Fe$_{1-x}$Mn$_{x}$\-As\-O$_{1-y}$\-F$_{y}$}

\newcommand{\zfmu}{ZF-$\mu$SR}
\newcommand{\tfmu}{TF-$\mu$SR}
\newcommand{\lfmu}{LF-$\mu$SR}
\newcommand{\musr}{$\mu$SR}

\begin{document}

\title{Role of magnetic dopants in the phase diagram of Sm1111 pnictides: The Mn case}

\author{G.~Lamura}
\affiliation{CNR-SPIN, Corso Perrone 24, I-161526 Genova, Italy}
\author{T.~Shiroka}
\affiliation{Laboratorium f\"ur Festk\"orperphysik, ETH-H\"onggerberg, CH-8093 Z\"urich, Switzerland}
\affiliation{Paul Scherrer Institut, CH-5232 Villigen PSI, Switzerland}
\author{S.~Bordignon}
\affiliation{Dipartimento di Fisica and Unit\`a CNISM di Parma, I-43124 Parma, Italy}
\author{S.~Sanna}
\affiliation{Dipartimento di Fisica and Unit\`a CNISM di Pavia, I-27100 Pavia, Italy}
\author{M.~Moroni}
\affiliation{Dipartimento di Fisica and Unit\`a CNISM di Pavia, I-27100 Pavia, Italy}
\author{R.~{De~Renzi}}
\affiliation{Dipartimento di Fisica and Unit\`a CNISM di Parma, I-43124 Parma, Italy}
\author{P.~Carretta}
\affiliation{Dipartimento di Fisica and Unit\`a CNISM di Pavia, I-27100 Pavia, Italy}
\author{P.~K.~Biswas}
\affiliation{Paul Scherrer Institut, CH-5232 Villigen PSI, Switzerland}
\affiliation{ISIS Pulsed Neutron and Muon Source, STFC Rutherford Appleton Laboratory,
Harwell Campus, Didcot, Oxfordshire, OX11 0QX, United Kingdom}
\author{F.~Caglieris}
\affiliation{CNR-SPIN, Corso Perrone 24, I-161526 Genova, Italy}
\affiliation{Dipartimento di Fisica, Universit\`a di Genova, via Dodecaneso 33, I-16146 Genova, Italy}
\author{M.~Putti}
\affiliation{CNR-SPIN, Corso Perrone 24, I-161526 Genova, Italy}
\affiliation{Dipartimento di Fisica, Universit\`a di Genova, via Dodecaneso 33, I-16146 Genova, Italy}
\author{S.~Wurmehl}
\affiliation{IFW Dresden, Institute for Solid State Research, P.O. Box 270116, D-01171 Dresden, Germany}
\author{S.~J.~Singh}
\affiliation{IFW Dresden, Institute for Solid State Research, P.O. Box 270116, D-01171 Dresden, Germany}
\affiliation{Department of Applied Chemistry, University of Tokyo, Japan}
\author{J.~Shimoyama}
\affiliation{Department of Applied Chemistry, University of Tokyo, Japan}
\author{M.~N.~Gastiasoro}
\affiliation{Niels Bohr Institute, University of Copenhagen, Juliane Maries Vej 30, 2100 Copenhagen, Denmark}
\author{B.~M.~Andersen}
\affiliation{Niels Bohr Institute, University of Copenhagen, Juliane Maries Vej 30, 2100 Copenhagen, Denmark}
\date{\today}

\begin{abstract}
The deliberate insertion of magnetic Mn dopants in the Fe sites of the optimally-doped Sm\-Fe\-As\-O$_{0.88}$\-F$_{0.12}$ iron-based superconductor can modify in a controlled way its electronic properties. The resulting phase diagram was investigated across a wide range of manganese contents ($x$) by means of muon-spin spectroscopy ($\mu$SR), both in zero- and in transverse fields, respectively, to probe the magnetic and the superconducting order. The pure superconducting phase (at $x<0.03$) is replaced by a crossover region at intermediate Mn values ($0.03 \leqslant x<0.08$), where superconductivity coexists with static magnetic order. After completely suppressing superconductivity for $x=0.08$, a further increase in Mn content reinforces the natural tendency towards antiferromagnetic correlations among the magnetic Mn ions. 
The sharp drop of $T_c$ and the induced magnetic order in the presence of magnetic disorder/dopants, such as Mn, are both consistent with a recent theoretical model of unconventional superconductors [M.\ Gastiasoro \textit{et al.}, ArXiv 1606.09495], which includes correlation-enhanced RKKY-couplings between the impurity moments.
\end{abstract}

\maketitle

\section{\label{sec:intro}Introduction}

The controlled insertion of disorder in superconducting (SC) materials via chemical substitution is a well known method to obtain valuable information regarding the gap symmetry. To this aim, several substitutions, either in the FeAs superconducting- or in the charge-reservoir layers of iron-based superconductors (IBS), have been regularly considered since their discovery in 2008.\cite{Hirschfeld2011} 
At the same time, this method demands particular caution in order to draw unambiguous conclusions about the gap symmetry, in particular in multiband systems.\cite{Hirschfeld2013,Gastiasoro2016} More generally, the use of diluted impurities represent a powerful tool for tuning the  superconductivity or inducing magnetic order (MO), with the detailed outcome depending on the nature of the impurity itself. Here we focus on the deliberate insertion of magnetic disorder in the FeAs layers of the 1111 family of superconducting compounds to evidence how and to what extent the presence of electronic correlations can enhance the magnetic coupling between diluted impurities and, hence, tune SC and MO, as suggested in a recent theoretical work.\cite{Gastiasoro2016}

Three types of Fe substitutions are possible: isovalent (Ru)\cite{Tropeano2010}, hole-dopant (Cr, Mn)\cite{Sato2012,Singh2013}, or electron-dopant (Co,Ni).\cite{Sato2012,Prando2013,Singh2013} In general, all of them induce a decrease of $T_c$, yet the decrease rate seems to depend significantly on the type of the substituted ion. The Mn-for-Fe substitution represents a particularly intriguing case, since even tiny amounts of manganese were shown to completely suppress the superconducting state.\cite{Sato2012} This type of substitution has been object of intense studies in the 122 IBS family, mostly because of the possibility to synthesize high-quality single crystals. In this case, magnetic resonance measurements (NMR and NQR) could show that, against the expections, Mn atoms do not induce any charge doping, most likely due to the localization of an additional manganese hole by the strong random potential induced by the magnetic Mn atoms.\cite{Bobroff2012} Photoemission and x-ray absorption measurements evidenced that large magnetic moments ($S = 5/2$) are formed at the Mn sites,\cite{Suzuki2013} around which small fluctuating regions with dominant nearest-neighbor AFM exchange interactions are created. The latter phase dominates at high Mn content as, e.g., in BaMn$_2$As$_2$ ($T_{\mathrm{N}} = 625$\,K),\cite{Kim2010,Tucker2012} where Mn ions play the role of \textit{localized magnetic scattering centers}. Interestingly, a new magnetic component, persisting well beyond the N\'eel temperature, sets in for Mn contents above a critical value\cite{Kim2010,Inosov2013}. Its presence was subsequently justified by taking into account
the magnetic coupling among the Mn impurities through the conduction electrons via the Ruderman-Kittel-Kasuya-Yosida interaction\cite{Gastiasoro2014}.

The role of Mn ions as localized magnetic impurities was confirmed by recent NMR-NQR studies also on 1111 compounds, such as \lfmao.\cite{Hammerath2014} The latter case seems particularly puzzling: very small amounts of Mn ($\sim 0.2$\%) are sufficient to fully suppress SC and to drive the system towards a short-range antiferromagnetic order. While passing through a quantum critical point at $x \sim 0.002$, the spin fluctuations progressively freeze and  competing low-frequency dynamic correlations (in the MHz range) develop as Mn content is increased.\cite{Hammerath2014,Hammerath2015}
Interestingly, the partial substitution of La with smaller Y ions drives the system away from quantum criticality, implying that a higher chemical pressure reduces the effects of Mn magnetic correlations.\cite{Moroni2016} It is conceivable that the same mechanism, i.e., a higher chemical pressure, reflecting the smaller size of Sm ions, can explain why $T_c$ decreases more slowly in the Sm-1111 than in La-1111 compounds for nominally equal Mn contents.\cite{Singh2013} Although a higher chemical pressure implies weaker electronic correlations, these still persist and have been shown to enhance the inter-impurity Ruderman–Kittel–Kasuya–Yosida (RKKY) exchange interaction, responsible for the competition between the magnetically-ordered and the superconducting phase.\cite{Gastiasoro2016}

In this paper, we investigate the evolution of the superconducting state and the type of magnetic correlations that develop in the \sfmaof\ system when the Mn content \textit{x} is systematically increased. Subsequently, we discuss our findings in the framework of the above mentioned theoretical work.\cite{Gastiasoro2016}

\begin{figure}[tbh]
\centering
\includegraphics[width=0.45\textwidth]{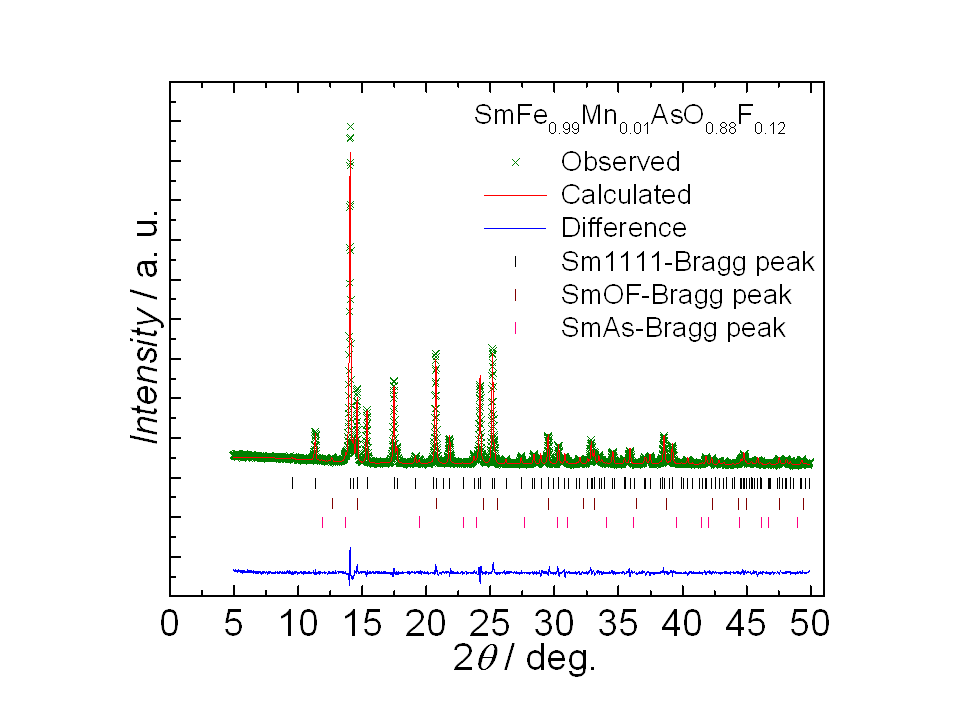} %rietveld.png xrd
\caption{\label{fig:rietveld}Typical x-ray diffraction pattern of an $x = 0.01$ sample and
the relevant Rietveld refinement. Notice the absence of spurious phases.}
\end{figure}
\begin{figure}[tbh]
\centering
\includegraphics[width=0.50\textwidth]{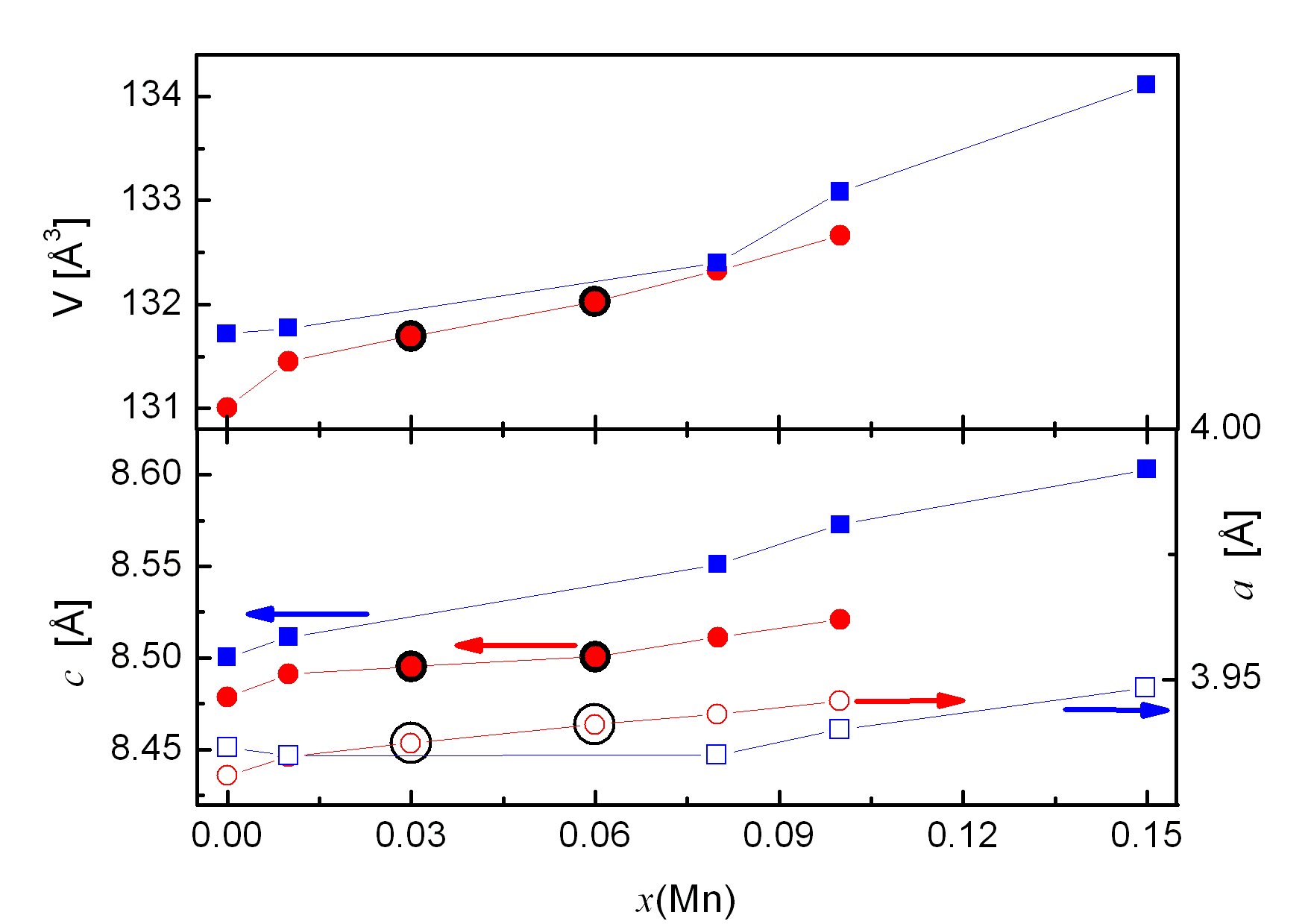} %latticeSmMn.png xrd
\caption{\label{fig:latticepar} Unit cell parameters vs.\ Mn content $x$ in \sfmaof\ : \textcolor{blue}{$\blacksquare$} refer to \#A series samples, \textcolor{red}{\Large $\bullet$} refer to \#B series, while {\Large \textbf{$\circ$}} highlight the two samples of \#B series used in this study. Void symbols in the bottom panel refer to $a$-axis values.}
\end{figure}
\begin{figure}[th]
\centering
\includegraphics[width=0.475\textwidth]{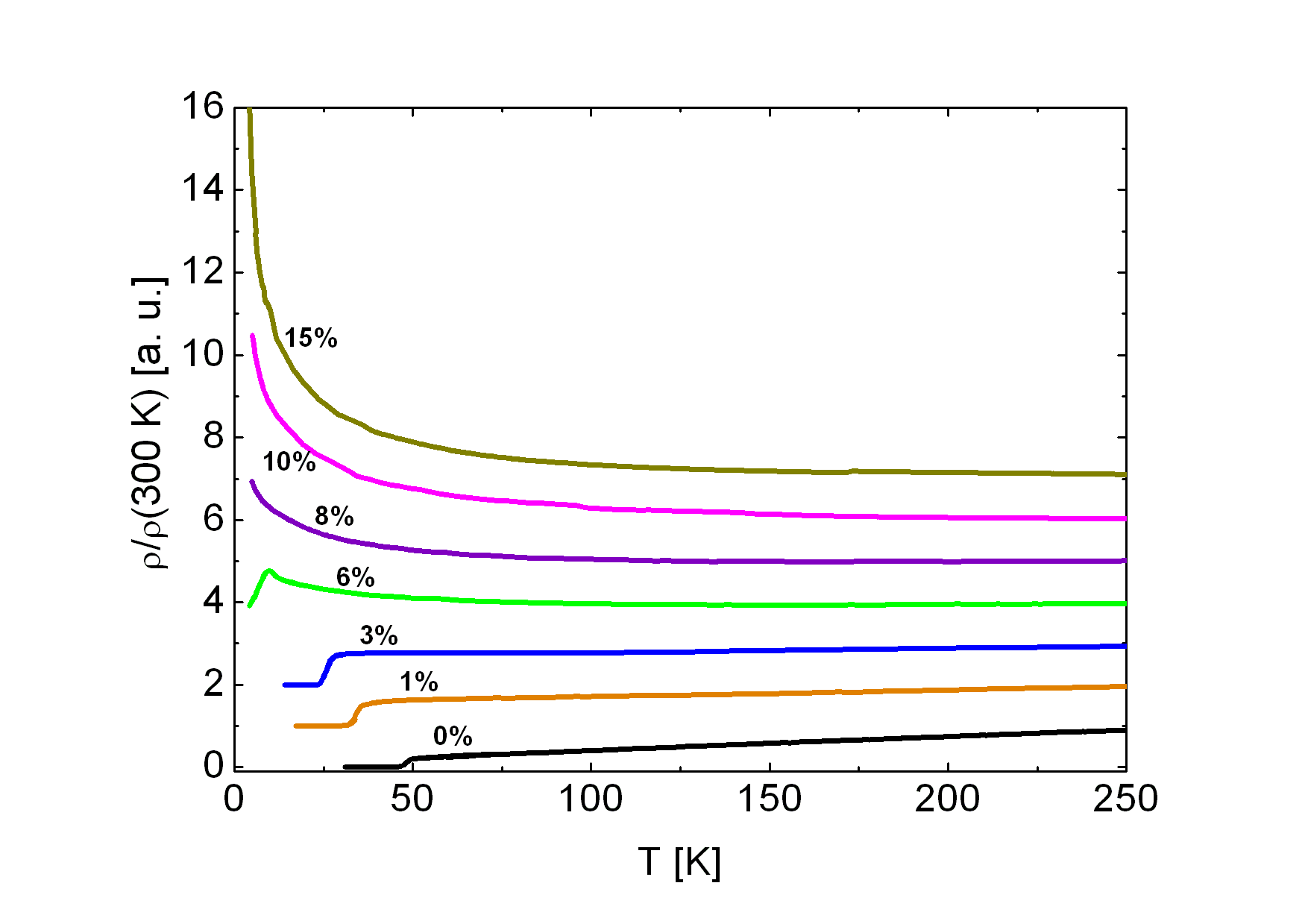} %resistivity.png
\caption{\label{fig:roALL}Normalized resistivity vs.\ temperature for \sfmaof\ samples. For the sake of clarity, each curve has been vertically shifted by one unit.}
\end{figure}
\section{\label{sec:preparation}Sample synthesis and characterization}
Two series of \sfmaof\ polycrystalline samples were investigated. The first one (\#A) was synthesized at IFW Dresden with nominal compositions $x = $ 0, 0.01, 0.08, 0.10, and 0.15. Another series (\#B) was prepared at the University of Tokyo with nominal compositions $x = $ 0, 0.01, 0.03, 0.06, and 0.1. Together they cover a broad range of Mn doping in finely-tuned steps. Samples of series \#A were synthesized in a single step, by using SmAs, Fe, Fe$_2$O$_3$, FeF$_3$, and Mn as precursors.
These were mixed according to the relevant stoichiometry and then thoroughly ground and pelletized. The pellets were inserted in quartz tubes which, after being evacuated and sealed, were heated in a furnace at 900$^{\circ}$C for 45\,h and then cooled down to room temperature at 150$^{\circ}$C/h. Samples of \#B series too were synthesized in a single-step solid-state reaction,\cite{Singh2013} but the last two precursors were FeF$_2$ and MnO (instead of FeF$_3$ and Mn). Compositions corresponding to different Mn doping values were mixed and pressed under 40 MPa into separate pellets, which were then wrapped in Ta foils and heated under identical conditions, following the same protocol as for the previous series.

Samples from both series were characterized by powder x-ray diffraction (XRD) using Mo-K$\alpha$ and Cu-K$\alpha$ radiation, respectively, with silicon powder being used as a standard reference. Rietveld analysis was employed to determine the lattice parameters of the \#A series, with an example for the $x=0.01$ case being shown in Fig.~\ref{fig:rietveld}. For the \#B series, instead, the lattice parameters were calculated based on the $d$ lattice spacing from the observed diffraction patterns  (see Ref.~\onlinecite{Singh2013}). For a comparison of the two series,  the cell parameter dependence vs.\ Mn content is shown in Fig.~\ref{fig:latticepar}. While the lattice parameter $a$ is mostly constant, the $c$-axis value increases linearly with increasing Mn content, with both sample sets showing a consistent behavior. As a result, by completing the series \#A with the $x = 0.03$ and 0.06 samples from series \#B, a full set of Mn-doping values could be achieved.
\begin{figure}[tbh]
\centering
\includegraphics[width=0.46\textwidth]{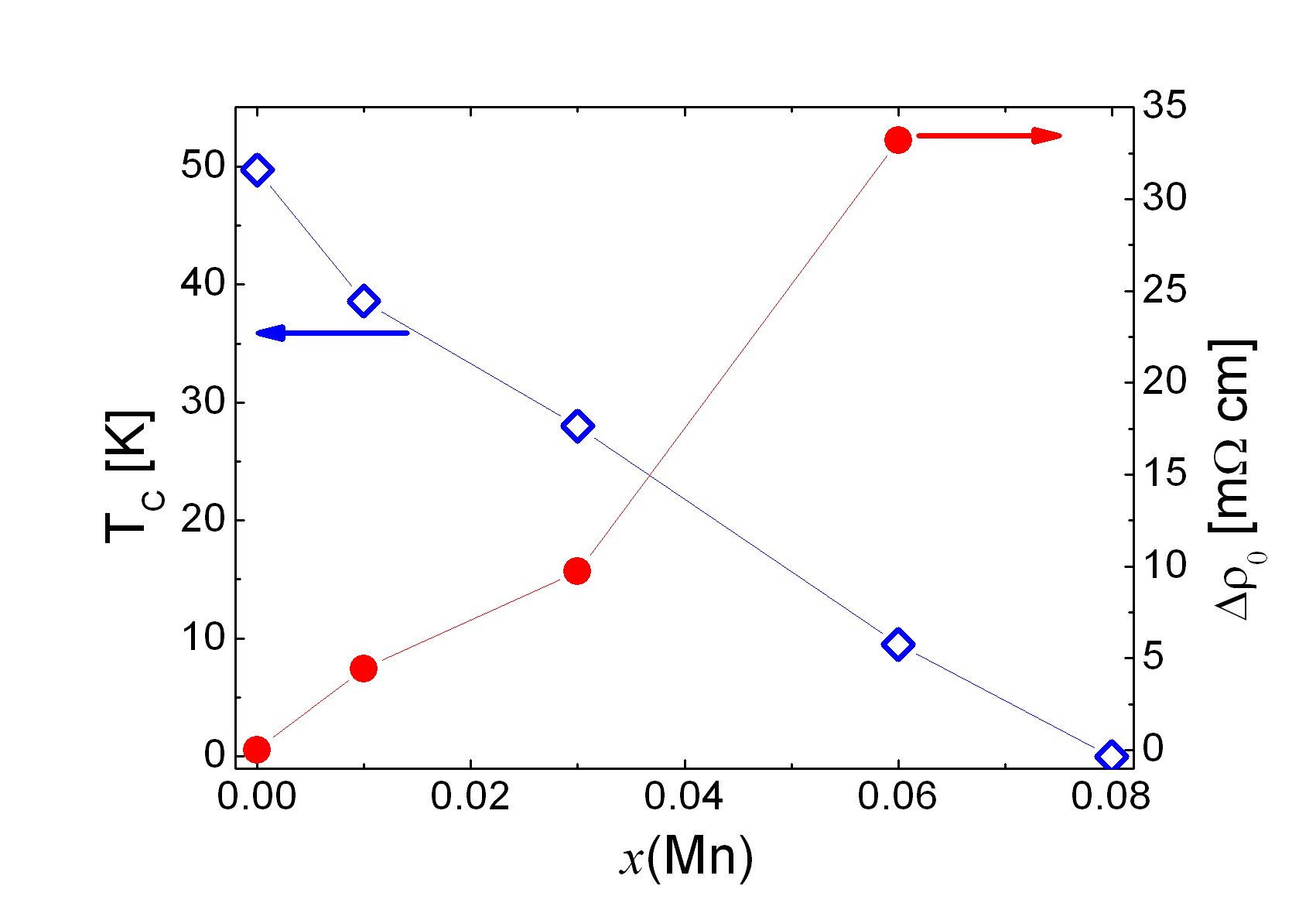} %rores.png ro residual-Tc vs. x(Mn)
\caption{\label{fig:rores} $T_{c}$ and excess resistivity $\Delta \rho_0$ vs.\ Mn content (see text for details).}
\end{figure}
\subsection{\label{ssec:transport}Resistivity measurements}
The resistivity of the \sfmaof\ samples was measured by means of a standard four-point method, with the temperature dependences $\rho(T)$ being shown in Fig.~\ref{fig:roALL}. Upon cooling, the resistivity of the $x=0$ sample decreases linearly with temperature down to the superconducting transition $T_{c} \simeq 46$\,K, defined as the feet of the resistivity drop. As the manganese content increases, the critical temperature decreases too, the superconductivity being fully suppressed for $x=0.08$. A further increase in Mn content results in an increase of the residual resistivity and in a low-temperature upturn, most likely due to weak localization effects. To evaluate the excess resistivity due to Mn dilution,
in the superconducting samples the high-temperature resistivity (i.e., from 100 to 250\,K) was fitted by a quadratic polynomial $\rho(x,T) = \rho_0(x)+a_1 T + a_2 T^2$, where $\rho_0(x)$ is the residual resistivity of the sample with a Mn fraction \textit{x}. The excess resistivity at $T=0$\,K was then estimated via $\Delta \rho_0 = \rho_0(x)-\rho_0$,\cite{Putti2010} with $\rho_0=0.20(2)$\,m$\Omega$\,cm the residual resistivity of the $x=0$ sample. Figure~\ref{fig:rores} shows the critical temperature $T_{c}$ and the excess resistivity $\Delta \rho_0$ vs.\ the Mn content: while $T_{c}$ decreases,  $\Delta \rho_0$ increases with $x$, both displaying an almost linear behavior.
The residual resistivities shown in Fig.~\ref{fig:rores} deserve some comment: (i) their magnitude is higher than expected due to grain-boundary effects reflecting the polycrystalline nature of our samples. By scaling the current resistivity data by a factor of ca.\ 1/4, one can remove the grain-boundary and anisotropy effects and hence estimate the in-plane resistivity.\cite{Sato2010} (ii) The low-$T$ resistivity values are generally lower than those reported in Ref.~\onlinecite{Luetkens2013}. For instance, the resistivity of the $x=0.10$ Sm-1111 sample is roughly 2/3 of the corresponding La-1111 sample. This suggests a lower degree of electronic correlation.\cite{Gastiasoro2016}
\begin{figure}[tbh]
\centering
\includegraphics[width=0.5\textwidth]{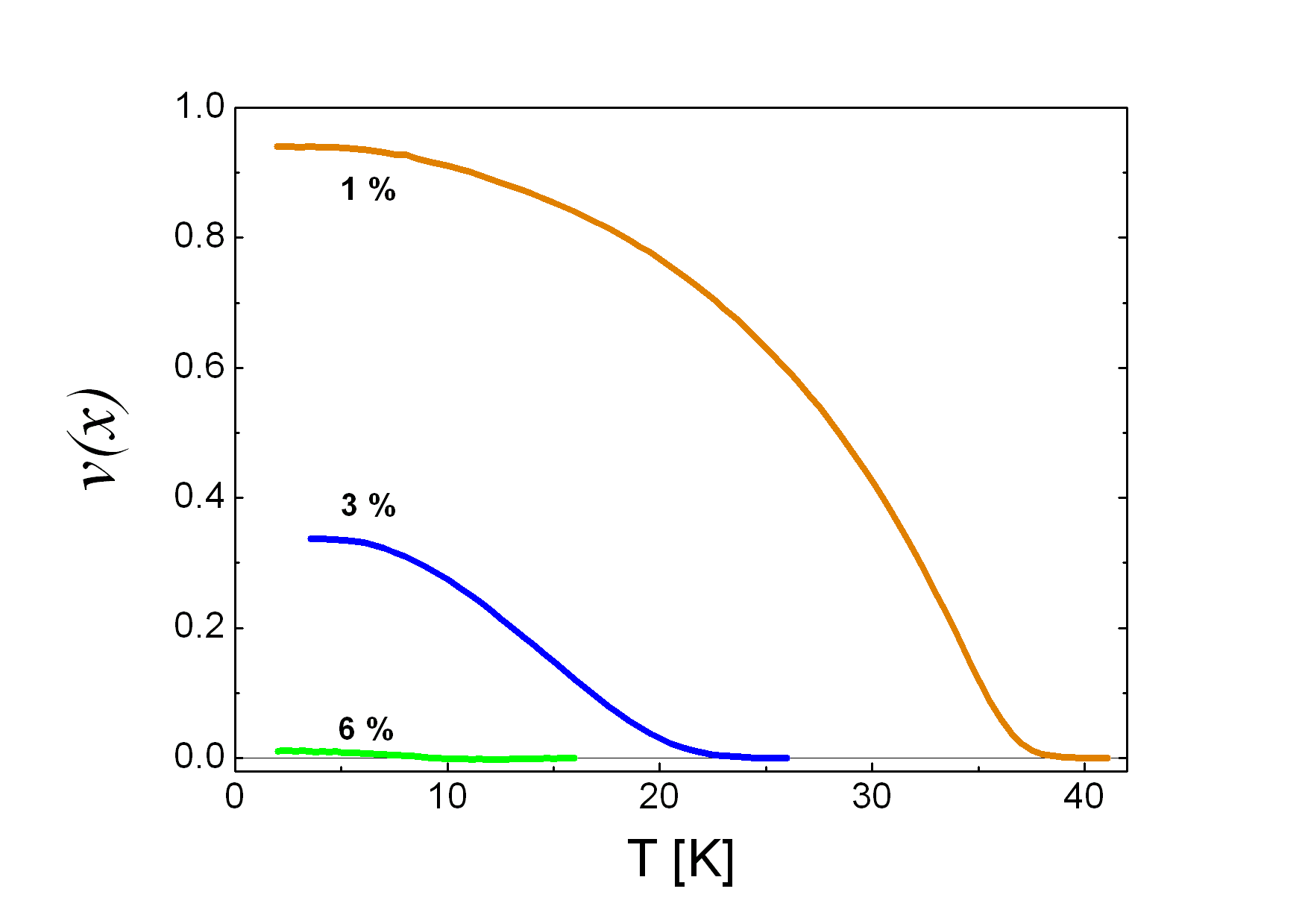} %MeissnerFraction.png nu(T) x all SC samples
\caption{\label{fig:MeissnerFraction}Superconducting volume fraction $\nu(T)$ for the superconducting samples (see text for details). }
\end{figure}

\subsection{\label{ssec:dcmagnetization}Magnetization measurements}
To characterize the superconducting state of the $x < 0.08$ \sfmaof\ samples, we carried
out dc susceptibility measurements on finely ground powders (demagnetizing factor $N=1/3$,
nominal density 7.5\,g/cm$^3$) by means of a superconducting quantum interference
device (SQUID) magnetometer. The susceptibility curves were measured at $\mu_0 H = 1$\,mT
from 2\,K to above $T_c$, both in zero-field-cooled (ZFC) and in field-cooled (FC) conditions.

To estimate the superconducting volume fraction we considered the variation of
FC susceptibility with respect to its normal-state value:
$\Delta \chi^{\mathrm{FC}}_{x}(T) = \left| \chi^{\mathrm{FC}}_{x}(T) - \chi^{\mathrm{FC}}_{x}(T>T_c) \right|$.
In the $x=1\%$ case, the superconductivity has a clear bulk character, as inferred
from TF-\musr\ data (see below), which determines a lower bound of $V_{sc}=94\%$
for the superconducting volume fraction.
Consequently, one can define the superconducting volume fraction $\nu(T)$ of
the other samples by normalizing their $\Delta \chi^{\mathrm{FC}}_{x}$ values to
$\Delta \chi^{\mathrm{FC}}_{[1\%]}(T_{\mathrm{min}})$:
\begin{equation}
\label{eq:V_SC}
\nu_{x}(T) = 0.94 \cdot \frac{\Delta \chi^{\mathrm{FC}}_{x}(T)}{\Delta \chi^{\mathrm{FC}}_{[1\%]}(T_{\mathrm{min}})}
\end{equation}
The resulting volume fractions $V_{sc}$, calculated by means of Eq.~{\ref{eq:V_SC}},
are shown in Fig.~\ref{fig:MeissnerFraction}. Clearly, to a moderate increase
in Mn content corresponds a steep decrease in $\nu(T=0)$.
\begin{figure}[th]
\centering
\includegraphics[width=0.5\textwidth]{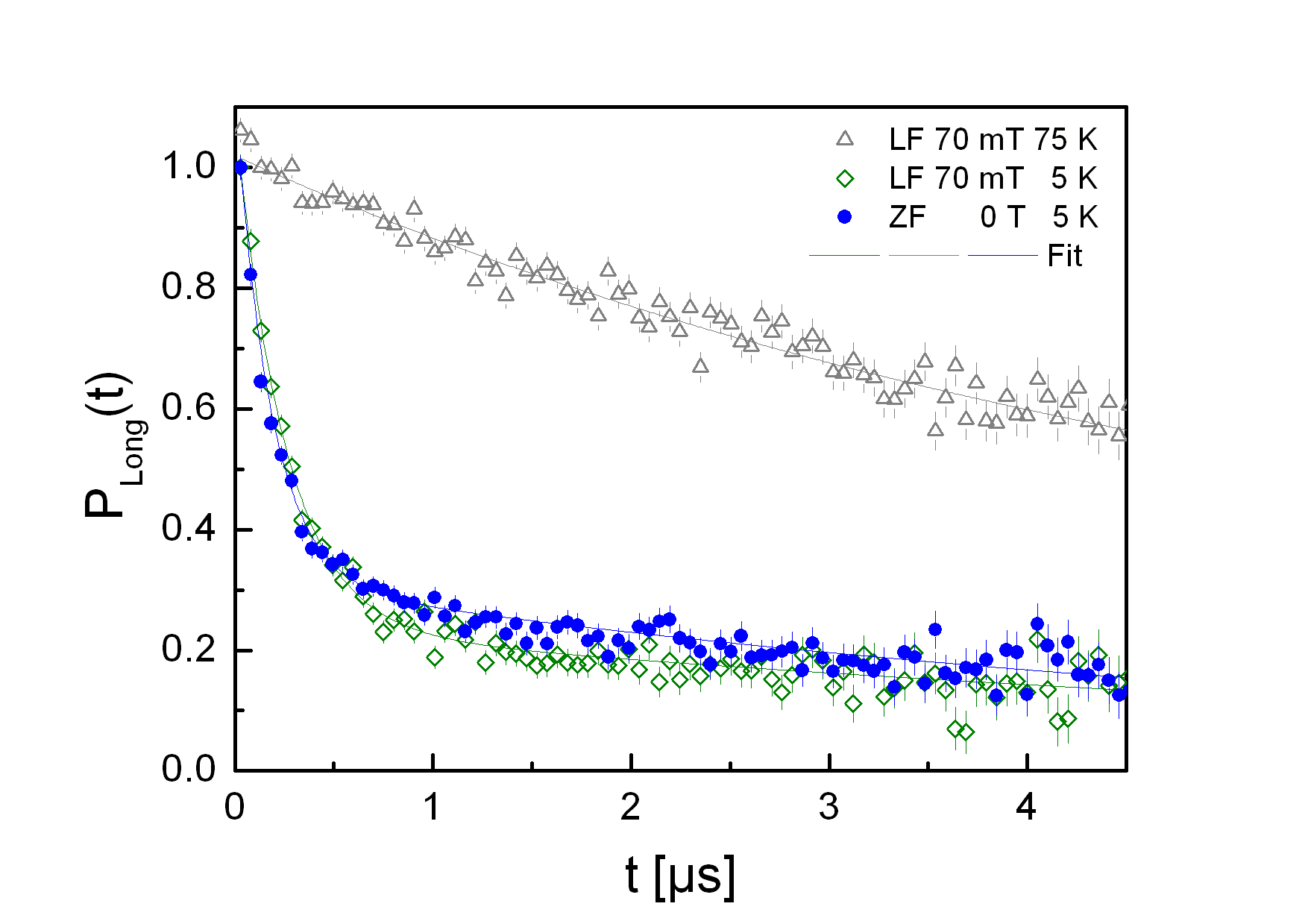} %pol_xMn01.png asymmetry_001
\caption{\label{fig:pol_xMn01}ZF- and LF-\musr\ short-time spectra for $x=0.01$
and $T=5$\,K at $\mu_0 H$=0\,T and $T=5$, 75\,K at $\mu_0 H$=70 mT (see text for details).}
\end{figure}
\begin{figure}[th]
\includegraphics[width=0.55\textwidth,angle=0]{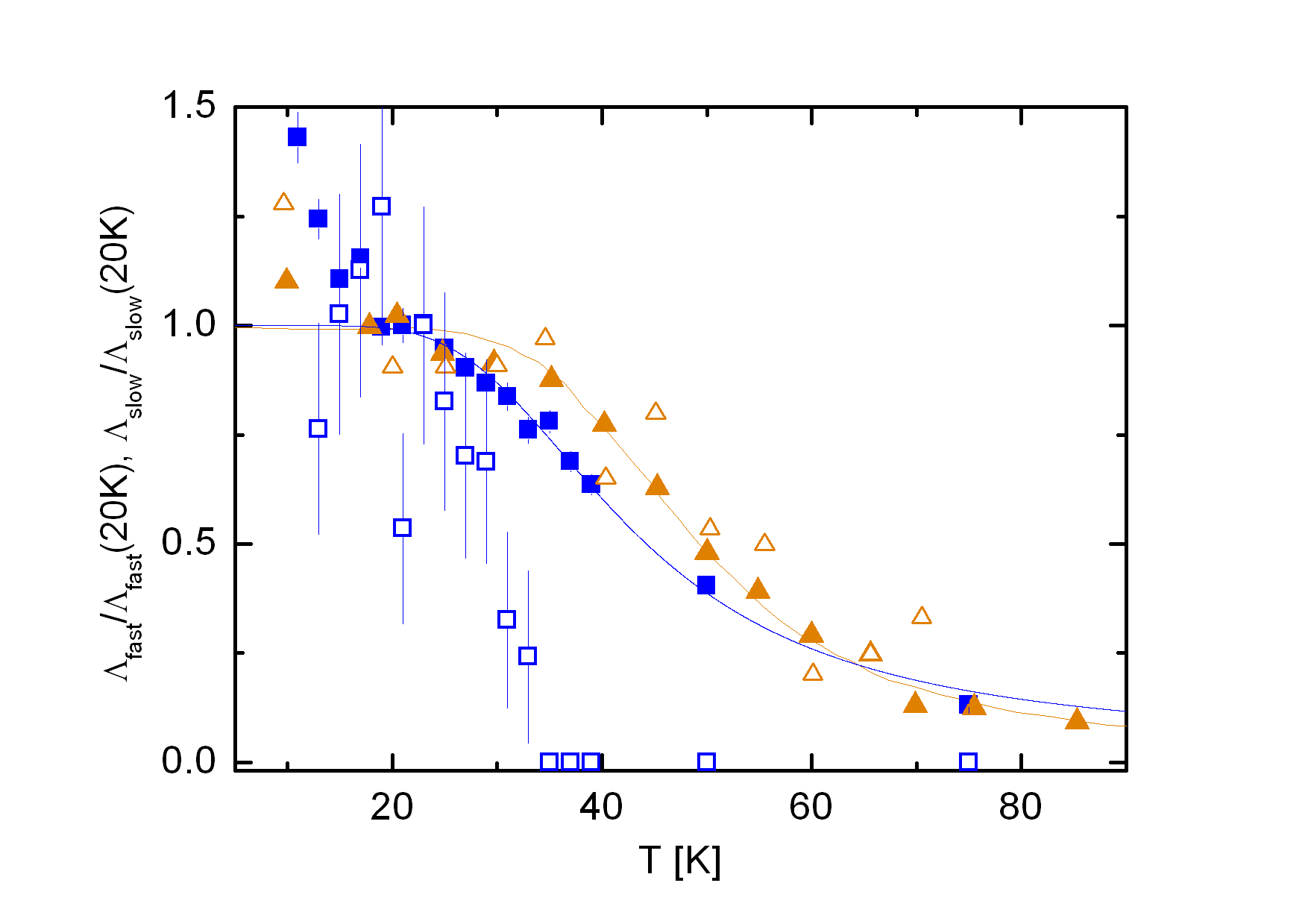} %LambdaRATIO.png
\caption{\label{fig:LambdaRATIO}a) Temperature dependence of $\Lambda_{\mathrm{fast}}/\Lambda_{\mathrm{fast}}(20\,\mathrm{K})$ (\textcolor{blue}{$\blacksquare$}) and $\Lambda_{\mathrm{slow}}/\Lambda_{\mathrm{slow}}(20\,\mathrm{K})$ (\textcolor{blue}{$\Box$}) as deduced from fits of \lfmu\ data by Eq.~(\ref{eq:P_LF}). Lines represent numerical fits by Eq.~(\ref{eq:Sm_level}). For a comparison, the normalized fast (\textcolor{orange}{$\blacktriangle$}) and slow (\textcolor{orange}{$\vartriangle$}) \musr\ relaxation rates, as measured in oxygen deficient Sm-1111 superconducting samples,\cite{Khasanov2008} are plotted too.}
\end{figure}

\section{\label{ssec:musr_exp}Muon-spin spectroscopy}
The muon-spin spectroscopy ($\mu$SR) measurements were carried out at the
GPS and Dolly instruments of the S$\mu$S facility at the Paul Scherrer Institute,
Switzerland.
Zero-field (ZF) $\mu$SR measurements were used to detect a possible spontaneous
magnetic order, or to distinguish between short- and long-range
order.\cite{Shiroka2011,Yaouanc2011} By means of longitudinal-field (LF) measurements
we could establish the static nature of magnetism (on the $\mu$SR timescale).\cite{Drew2008,Yaouanc2011}
Finally,  by transverse-field (TF)  $\mu$SR experiments we could determine the properties
of the vortex lattice in the superconducting phase.
The relatively large samples' thickness (about 1\,mm) and the use of veto counters implied good signal-to-noise ratios,
hence ensuring that in all the experiments the signals were due exclusively to muons stopped in the samples.

\begin{figure}[ht]
\includegraphics[width=0.5\textwidth,angle=0]{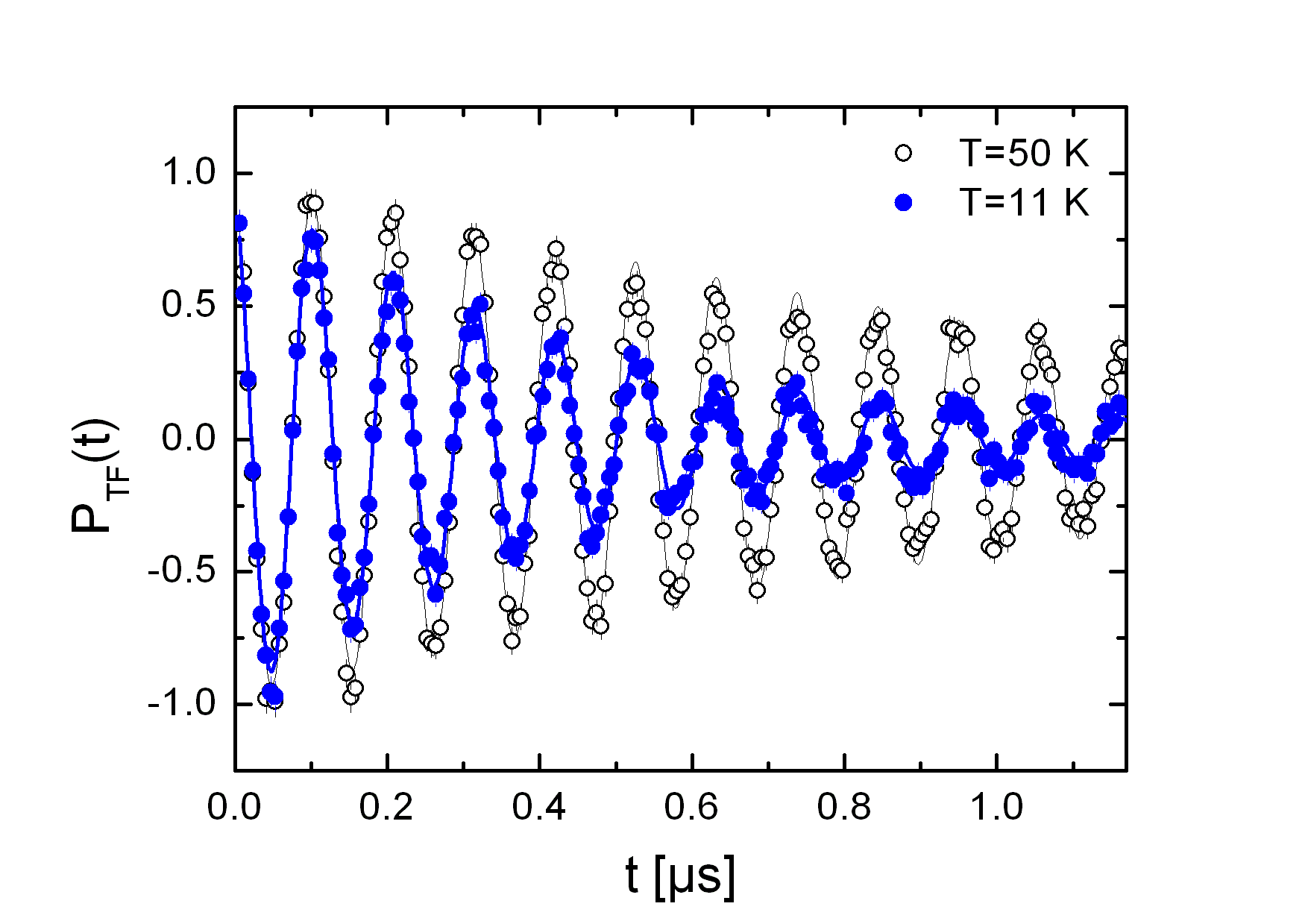} %polSC.png
\caption{\label{fig:polSC}(Color online) Time-dependent \tfmu\ polarization for the $x = 0.01$ sample measured at $T=11$ and 50\,K.}
\end{figure}
\subsection{\label{ssec:h001}Superconducting properties as probed by \musr}
\subsubsection{Fluctuating magnetism due to paramagnetic Sm ions}
Figure~\ref{fig:pol_xMn01} shows the low-temperature ZF muon-spin polarization 
$P(t)=A(t)/A(0)$, with $A(t)$ and $A(0)$ being the time-dependent and initial asymmetry, 
respectively. It is relevant to note that the significant depolarization observed at short times 
($t < 1$\,$\mu$s) is insensitive to the application of a 70-mT longitudinal field, yet 
the depolarization is markedly reduced at $T = 75$\,K,  well above the superconducting critical temperature. 
Both ZF and LF $P(t)$ data could be fitted by means of:\cite{Khasanov2008}
\begin{equation}
\label{eq:P_LF}
P_{\mathrm{ZF},\mathrm{LF}}(t)=  p_{\mathrm{fast}} e^{-\Lambda_{\mathrm{fast}} t} + p_{\mathrm{slow}} e^{-\Lambda_{\mathrm{slow}} t},
\end{equation}
where $p_{\mathrm{fast}/\mathrm{slow}}$ and $\Lambda_{\mathrm{fast}/\mathrm{slow}}$ are the relative weights and relaxation rates of muons implanted in two inequivalent sites, namely close to the SmO and to the FeAs planes.\cite{Khasanov2008,Drew2008,Sanna2009,DFT_2012,DFT_2013,DFT_2015} The different relaxation rates reflect the different dipolar fields created by the Sm${}^{3+}$ moments at the two muon sites. LF data were fitted by keeping the $p_{\mathrm{fast}}/p_{\mathrm{slow}}$ ratio fixed. At $T \gtrsim 75$\,K, the muon polarization could be adequately fitted by a single-exponential decay.
\begin{figure}[th]
\includegraphics[width=0.45\textwidth,angle=0]{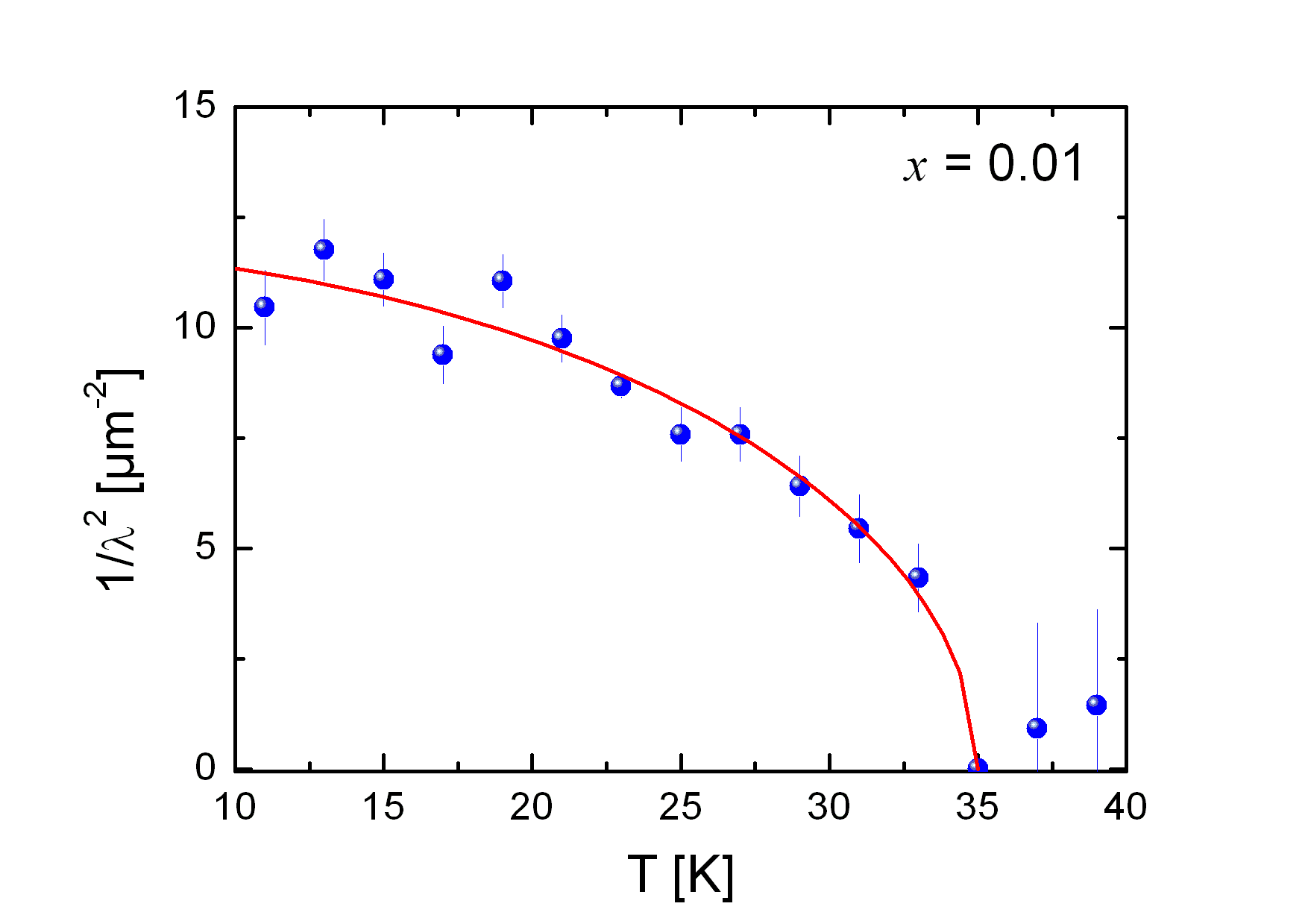} %sigmaSC_2.png
\caption{\label{fig:sigmaSC_2}(Color online) Temperature dependence of $\lambda^{-2}$ of the $x=0.01$ sample reconstructed from $\sigma_{sc}$ measured in $\mu_0 H = 70 mT$. The curve represents a fit using an average-field model. Data are limited to 10\,K, since at lower temperatures depolarization effects due to Sm paramagnetism prevent the extraction of $\lambda$.}
\end{figure}
Figure~\ref{fig:LambdaRATIO} shows the
temperature variation of the normalized $\Lambda_{\mathrm{fast}}$
and $\Lambda_{\mathrm{slow}}$ values. While the former agrees well with previous 
results on superconducting Sm-1111 samples,\cite{Khasanov2008,Drew2008}  
the latter shows a moderate agreement only at low temperature. 
At high temperature, instead, muon diffusion apparently narrows 
the lineshape, with the resulting slow relaxation becoming too small to be detected. 
Consequently, we limit our discussion to the fast relaxation term only. 
Generally, the observed behavior was ascribed to Sm${}^{3+}$ magnetic moment fluctuations. 
In a simplified model, the Sm-moment
fluctuation rate $1/\tau_c$ can be considered as the sum of a $T$-independent 
and a thermally-activated term.\cite{Carretta2009} The latter accounts for 
the thermally populated Sm crystal-field levels via:\cite{Khasanov2008,Cimberle2009}
\begin{equation}
\label{eq:Sm_level}
\frac{1}{\tau_c}=\frac{1}{\tau_c(0)}+\frac{1}{C e^{E_0 / k_\mathrm{B} T}},
\end{equation}
where $E_0$ is an activation energy and $C$ is a constant.
In the fast fluctuation limit the measured relaxation rate is $\Lambda=(\gamma_\mu^2 \langle \Delta h^2_\perp
\rangle) \cdot \tau_c$, with $\langle \Delta h^2_\perp \rangle$
the mean-square amplitude of the fluctuating field 
generated by the Sm$^{3+}$ moments at the muon sites.\cite{Carretta2009} 
Since for $k_\mathrm{B}T\ll E_0$ the system is in its ground-state, 
the field amplitude due to the fluctuating Sm${}^{3+}$ ions can be considered 
as temperature independent. Therefore, the only temperature dependence of
$\Lambda$ can come from $\tau_c$. Accordingly, the
relaxation rate assumes the empirical form already used for
SmFeAsO (see, e.g., Refs.~\onlinecite{Khasanov2008,Maeter2009}). 
To verify how and to what extent the Sm moment fluctuations are altered by 
the Mn dilution, we fit the $\Lambda_\mathrm{fast}(T)$ data 
by using the $\tau_c(T)$ dependence as given by Eq.~\ref{eq:Sm_level}. 
For the fits, only the $\Lambda_\mathrm{fast}(T)$ data in the 20--100\,K temperature 
range were considered, so as to exclude the low-temperature upturn due to the 
magnetic ordering of the Sm sub-lattice. 
The fit (see Fig.~\ref{fig:LambdaRATIO}) yields $E_0=15\pm1$\,meV, close to the value reported in Ref.~\onlinecite{Khasanov2008}. The
analogous temperature dependence of the measured relaxation rates,
including their slowing down around 80\,K, and their insensitivity
to the superconducting phase, confirms that these fluctuations are
due to the Sm paramagnetic moments. A slight reduction in the
excitation-gap value with respect to the known
values,\cite{Khasanov2008,Cimberle2009} could be ascribed to the
influence of the diluted Mn magnetic moments.
\subsubsection{Superconducting properties}
\begin{figure}[th]
\includegraphics[width=0.4\textwidth,angle=0]{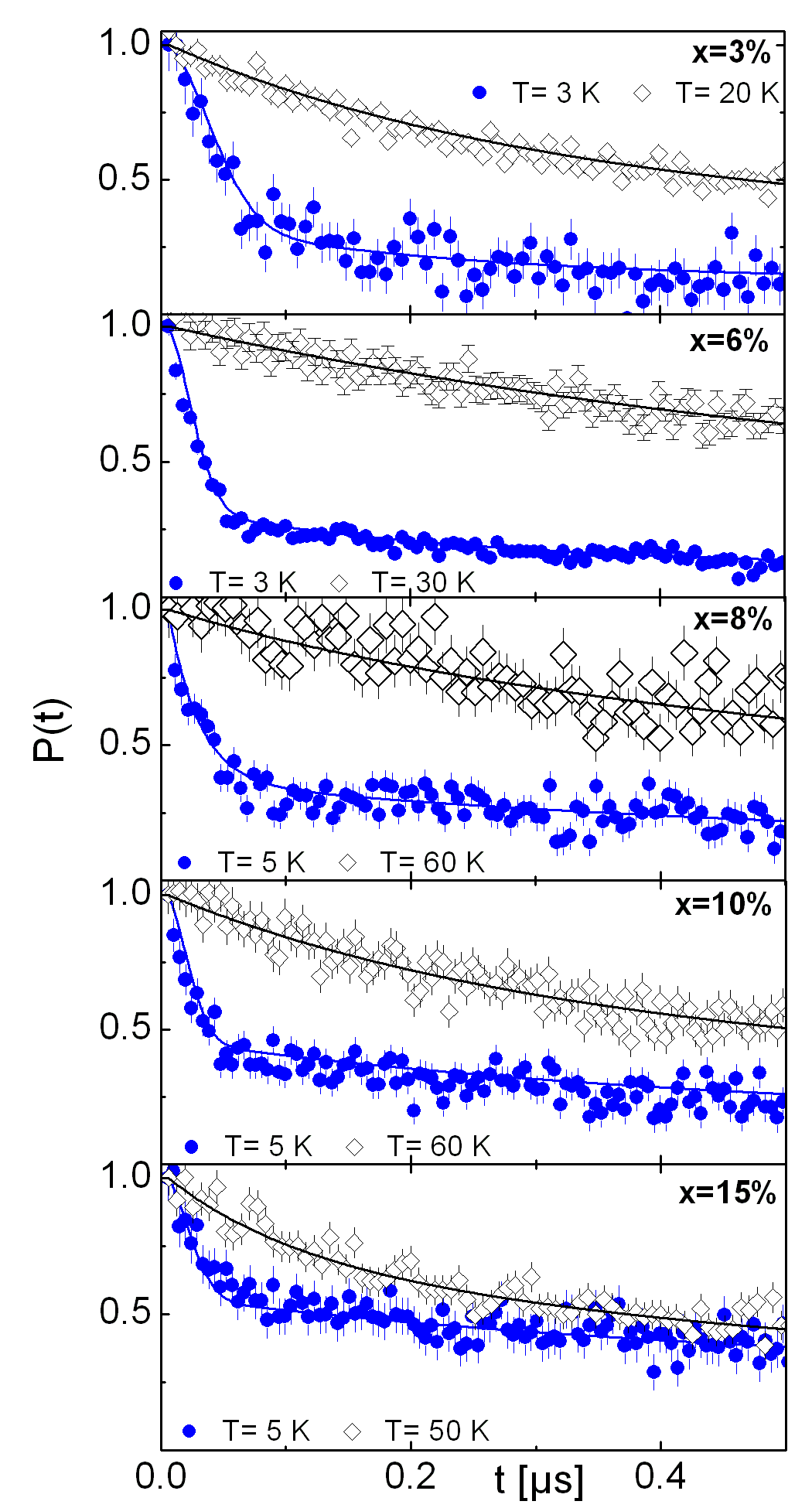} % asymmetryALL.png
\caption{\label{fig:asymmetryALL}(Color online) ZF-$\mu$SR short-time spectra of magnetic \sfmaof\ samples at selected temperatures. The significant increase of damping below 60\,K suggests the onset of a magnetically ordered phase.The continuous lines represent numerical fit eq.~\ref{eq:spin_prec}.}
\end{figure}
TF-$\mu$SR spectroscopy measurements were performed on the $x=0.01$, 0.03, and 0.06 samples. Unfortunately, the presence of static antiferromagnetic order in the FeAs planes (see Section III.B), coexisting with superconductivity, prevented a reliable evaluation of the magnetic penetration depth in the $x=0.03$ and 0.06 case. However, since no traces of static magnetism could be found down to 1.5\,K in the $x=0.01$ sample, its superconducting properties could be investigated in detail via TF-$\mu$SR after cooling the sample in an applied field $B_0 = 70$\,mT.
Figure~\ref{fig:polSC} shows the short-time polarization of the muon ensemble below and above $T_c$. To disentangle the depolarization effects due to the vortex lattice from contributions due to the fluctuating Sm paramagnetic moments, the time-dependent transverse polarization in the SC phase was fitted by means of the equation:\cite{Khasanov2008}
\begin{equation}
\label{eq:P_TF}
P_{\mathrm{TF}}(t)=P_{\mathrm{LF}}(t) e^{-\frac{\sigma t^2}{2}}  \cos(\gamma_{\mu} B_{\mu} t + \phi) + b(t),
\end{equation}
where $P_{\mathrm{LF}}(t)$ is defined in Eq.~(\ref{eq:P_LF}), $B_{\mu}$ is the average field at the muon site, $\gamma_{\mu} = 2\pi \times 135.53$\,MHz/T is the muon gyromagnetic ratio, $\phi$ the initial phase, $\sigma$ the Gaussian relaxation rate, and $b(t)$ represents the normal-phase signal, whose relaxation and oscillating frequency (exactly at $B_0$) were determined above $T_c$ and then kept fixed. 
The relative weight of $b$ was determined at long times ($t \gg 1/\sigma$) and the lowest temperature: in the superconducting state it represents only $\sim$6.4\% of the total signal, which implies a Meissner fraction of about 94\%.

At each temperature, the parameters defining the $P_{\mathrm{LF}}(t)$ term were taken from the corresponding LF parameters at the same temperature [see Eq.~(\ref{eq:P_LF})].

As for the Gaussian depolarization rate, it consists of a superconducting contribution ($\sigma_{sc}$), a magnetic contribution due to the ordering of Sm ions ($\sigma_{m}$), and a nuclear magnetic contribution ($\sigma_{nm}$), the latter being determined in the normal state:\cite{Khasanov2008,Drew2008}
\begin{equation}
\label{eq:sigma}
\sigma^2=\sigma^2_{sc}+\sigma^2_{m}+\sigma^2_{nm}.
\end{equation}

Since the Sm contribution is relevant only at low temperatures, by considering the data above 10\,K the $\sigma_{m}$ term in Eq.~(\ref{eq:sigma}) is negligible, hence enabling us to extract $\sigma^2_{sc}$. For anisotropic superconductors in the limit of low fields, the effective magnetic penetration depth $\lambda$ is related to $\sigma_{sc}$ through the equation:\cite{Brandt1988}
\begin{equation}
\label{eq:lambda}
\frac{\sigma_{sc}^2}{\gamma^2_{\mu}}=0.00371 \cdot \frac{\varPhi_0^2}{\lambda^4},
\end{equation}
%
%1 Wb = 1 Tm2  1 um2 = 10-12 m2
where $\varPhi_0 =2.068\times10^{-3}$\,T$\mu$m$^2$ is the quantum of magnetic flux. Figure~\ref{fig:sigmaSC_2} shows the temperature dependence of the superfluid density, $n_s \propto \lambda^{-2}(T)$, and a numerical fit with an average-field  model $1/\lambda^2(T)=(1/\lambda^2(0))[1-(T/T_c)^n]$, which gives $1/\lambda^2(0)=11.9\pm0.7$\,$\mu$m$^{-2}$ and $n=2.0\pm0.4$. Subsequently, by considering that in anisotropic polycrystalline samples the relation $\lambda_{\mathrm{eff}}(0)= 3^{1/4}\lambda_{ab}(0)$ holds,\cite{Fesenko1991} we could estimate the in-plane magnetic penetration depth $\lambda_{ab}(0) = 221\pm7$\,nm. Since this value is slightly higher than that measured in optimally doped Sm-1111 samples,\cite{Khasanov2008,Drew2008} it suggests a reduced zero-temperature superfluid density in the Mn-doped case.
\begin{figure}[ht]
\includegraphics[width=0.40\textwidth,angle=0]{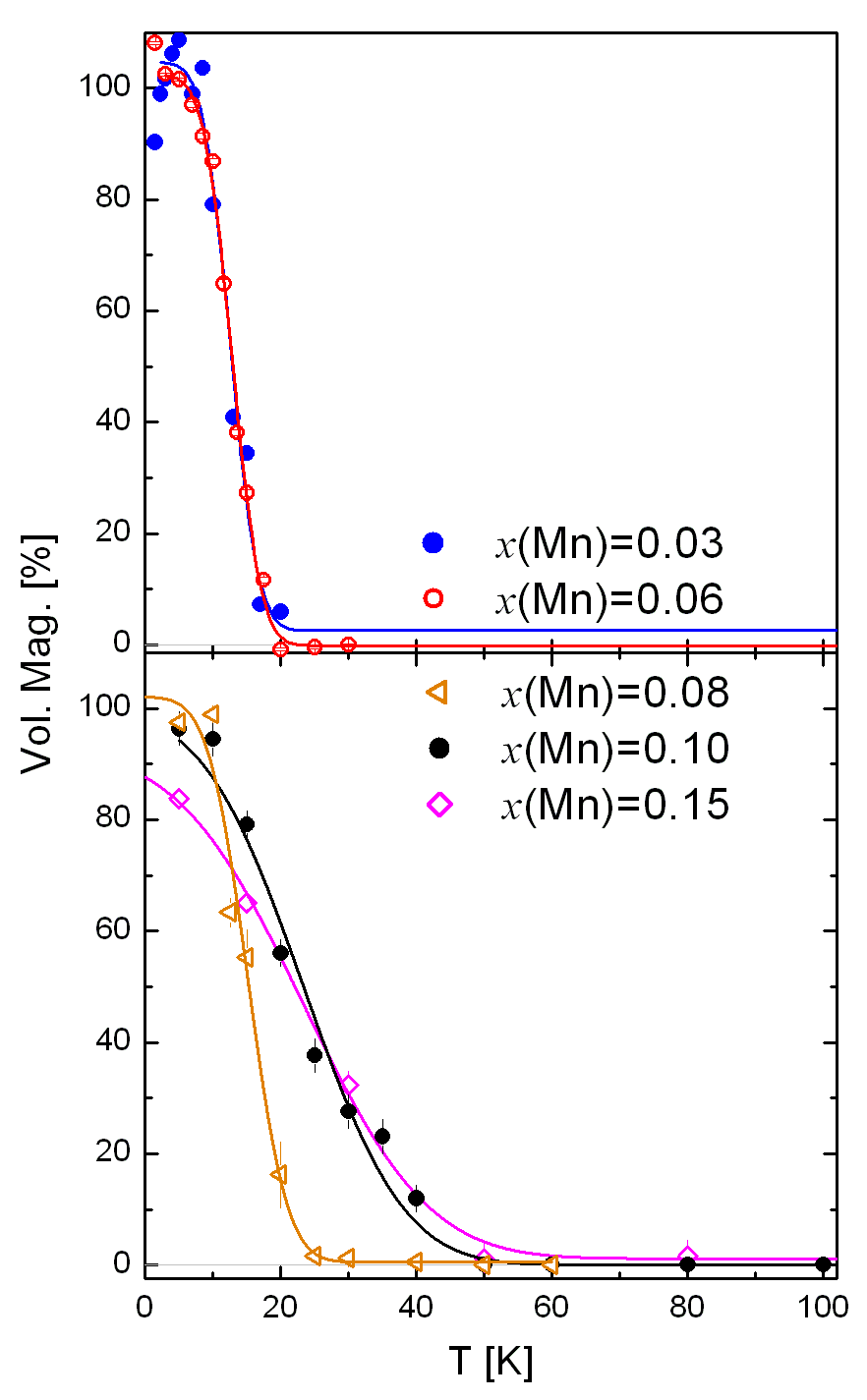} % vol_mag_last.png
\caption{\label{fig:Vm_ALL}(Color online) Temperature dependence of the magnetic volume fractions for the $x=0.03$, 0.06 (top panel) and $x=$ 0.08, 0.10, 0.15 samples (bottom panel), respectively. The lines are numerical fit by means of an \textit{erf} function.}
\end{figure}
\subsection{ZF-$\mu$SR and magnetic properties}
Figure~\ref{fig:asymmetryALL} shows a selection of ZF spectra for all those \sfmaof\ samples where we found evidence of static magnetism. The most prominent features are the presence of a rather large relaxation at high temperatures and a significantly damped signal with no coherent precessions below about 60\,K.
The time-dependent polarization was fitted by the following model:
\begin{multline}
%\begin{equation}
\label{eq:spin_prec}
P_{\mathrm{ZF}}(t) =  \left[1-V_M(T)\right] \cdot g(t)+\\
+ V_M(T) \sum_{i=1}^{N} w_i \cdot \left[p_{T_i} \,
f_i(\gamma_{\mu}B_{\mu} t) \, D_{T_i}(t) + p_{L_i} \, D_{L_i}(t) \right],
%\end{equation}
\end{multline}
where $V_M$ is the magnetic volume fraction and $g(t)$ the time-dependent relaxation in the paramagnetic state. In the magnetically ordered state a nonzero $V_M$ fraction of muons probes a local magnetic field $B_\mu$ at the implantation site $i$; $p_{T_i}$ and $p_{L_i}$ in Eq.~(\ref{eq:spin_prec}) refer to muons probing local fields in the transverse (T) or longitudinal (L) directions with respect to the initial muon-spin polarization. The coherent precession of muons is taken into account by the $f(t)$ function, whereas $D_{T_i}(t)$ and $D_{L_i}(t)$ model the precession damping. The decay $D_{T_i}(t)$ reflects the static distribution of local magnetic fields, whereas $D_{L_i}(t)$ is due to dynamical relaxation processes. Finally, the sum over \textit{i} generalizes Eq.~(\ref{eq:spin_prec}) to the case of several inequivalent crystallographic implantation sites, whose relative populations $w_i$
are normalized to 1.
\begin{figure}[ht]
\centering
\includegraphics[width=0.5\textwidth]{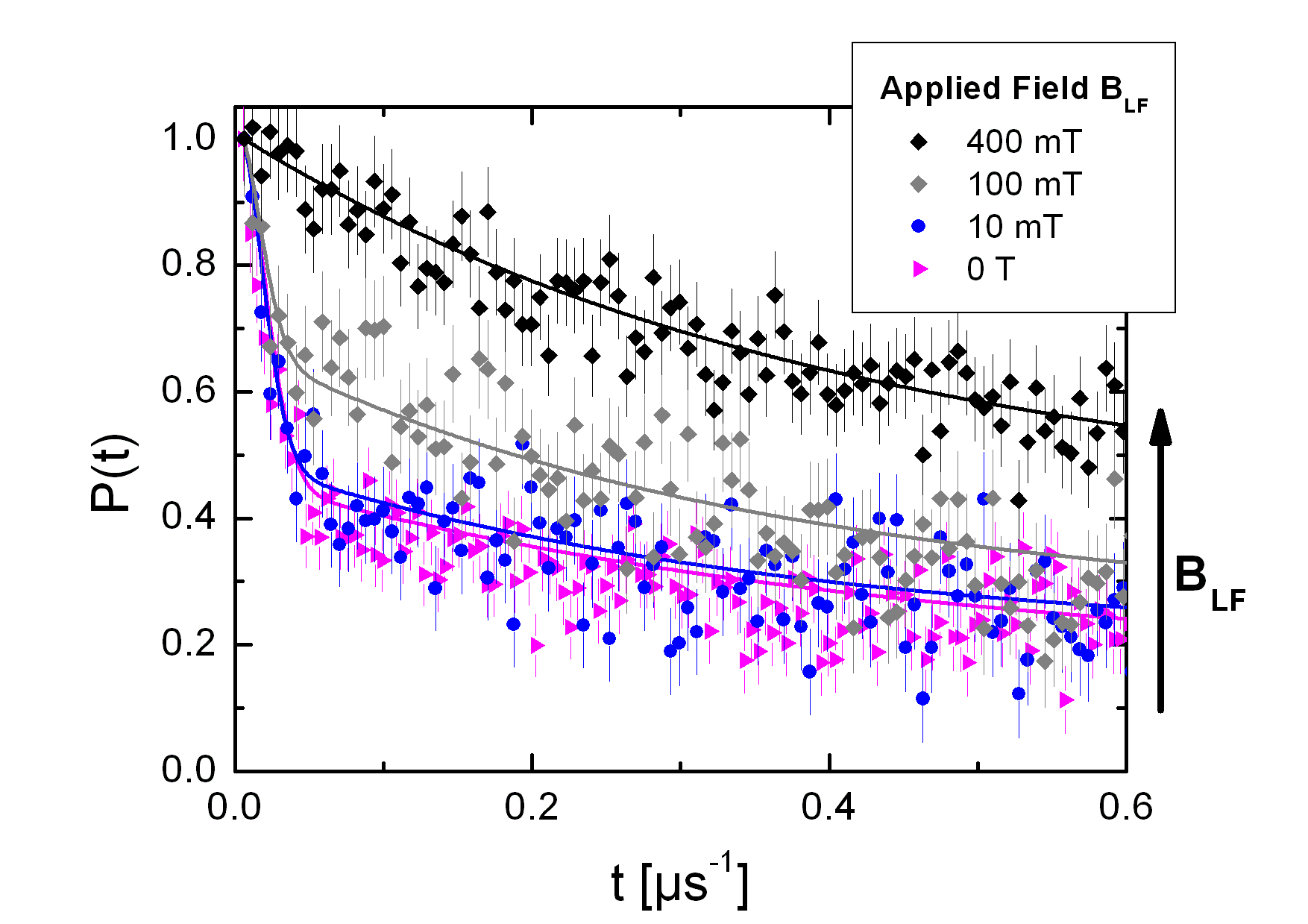} % asymmetryLF10.png  %{LF_01} %LFscan
\caption{\label{fig:asymmetryLF10}(Color online) LF-$\mu$SR time-domain spectra measured
at $T = 5$~K in the $x=0.10$ sample. The continuous lines represent numerical fits using Eq.~(\ref{eq:P_LF}) (see text for details).}
\end{figure}
Yet, Sm-1111 is a peculiar pnictide system, since the implanted muons in the FeAs and SmO planes couple differently with 
the Sm magnetic moment fluctuations, always detectable on the \musr\ time scale\cite{Drew2008,Khasanov2008,Sanna2009,Drew2009,Maeter2009}. For this reason, differently from other SC pnictides, the longitudinal relaxations arising from muons implanted in two different sites are distinct in our case in both the paramagnetic and the magnetically ordered state.

In the magnetically ordered phase ($T <  T_\mathrm{N}$), a very strong damping below about 60\,K
indicates the onset of magnetic order. For $x>0.03$ the best fits were obtained by reducing Eq.~(\ref{eq:spin_prec}) to 
the sum of a transverse component (with Gaussian decay) and two longitudinal ones, corresponding to the longitudinal relaxations (fast/slow) discussed above.\cite{Drew2009} Only for the $x=0.03$ sample the magnetically ordered phase was fitted by one transverse and one longitudinal exponential terms.
In the high-temperature paramagnetic phase ($T> T_\mathrm{N}$) some small differences arise: for $0.03\leq x \leq0.06$ the best fit was obtained by using two Lorentzian relaxation terms, as described by Eq.~(\ref{eq:P_LF}), which suggests the presence of fast fluctuating magnetic moments. For $x>0.06$, instead, the $g(t)$ term is best described by the sum of two Lorentzian Kubo-Toyabe (KT) functions, more suitable for fitting large relaxation rates
arising from homogeneously diluted ferromagnetic impurities.\cite{Sanna2009JSNM,Lamura2014}

In Table~\ref{tab:TCTN} we summarize the internal magnetic field widths $\Delta B_{\mu}$ at 5\,K, as
determined from the Gaussian decay of the transverse component.
Figure~\ref{fig:Vm_ALL}, instead, shows the temperature dependence of the magnetic volume fraction $V_M$ of the magnetically-ordered
phase, calculated from the total longitudinal component using $V_M(T) = \frac{3}{2} \left(1-a_{\parallel}\right)$.\footnote{In magnetically ordered polycrystalline samples with domains distributed isotropically overall solid angles, 1/3 of the implanted muons probe a local field parallel to their initial polarization, whereas for 2/3 the initial polarization is orthogonal to it.}
In Table~\ref{tab:TCTN} we report the magnetic transition temperatures and their widths, as obtained
by a phenomenological \textit{erf}-like fit of $V_M(T)$ data. All the samples with $x\geq0.03$ share the following features: (i) they
are fully magnetically ordered at low temperature; (ii) both the magnetic transition width and the average N\'eel temperatures increase with increasing Mn content. Interestingly, the increased broadening of the magnetic transition mimics the behavior evidenced in 122 systems: for Mn concentrations above a critical threshold a new magnetic component (under the form of a long tail), persisting well beyond $T_\mathrm{N}$, appears in the magnetically ordered phase\cite{Kim2010,Inosov2013}. This component was ascribed to the magnetic coupling of Mn ions by conduction electrons via the RKKY interaction\cite{Gastiasoro2014}.
\begin{figure}[ht]
\includegraphics[width=0.5\textwidth,angle=0]{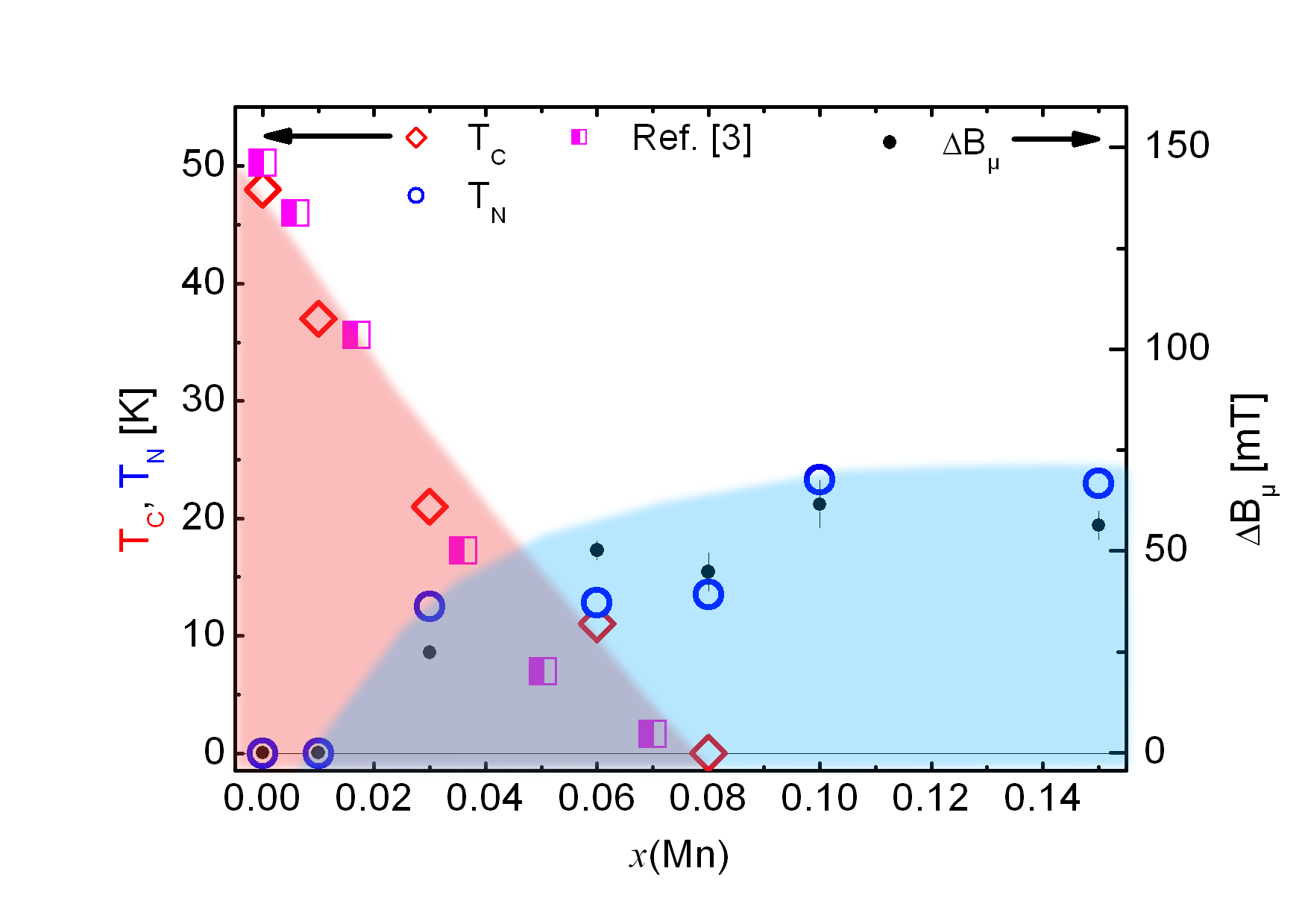} % PhaseDiag8.png
\caption{\label{fig:PhaseDiag}(Color online) Magnetic ordering $T_{\mathrm{N}}$ (\textcolor{blue}{$\bullet$}) and critical superconducting temperature  $T_c$ (\textcolor{red}{\large{$\diamond$}}) vs.\ Mn content $x$ in the \sfmaof\ family. The internal field widths $\Delta B_{\mu}$ (\textbf{$\bullet$}) were determined from the decay of the transverse Gaussian component by fitting the ZF-$\mu$SR data with Eq.~(\ref{eq:spin_prec}). Squares represent theoretical calculations from Ref.~\onlinecite{Gastiasoro2016}. }
\end{figure}

In principle, the absence of a coherent muon precession could be due either to a wide distribution of static fields, or to strongly fluctuating (i.e., dynamic) magnetic moments. To check if the magnetically ordered phase evidenced by \zfmu\ is static in nature (on the \musr\ time-scale), we carried out an LF-decoupling experiment (\lfmu\ ) in the representative $x=0.10$ case. In \lfmu\ experiments an external magnetic field $B_{\parallel}$ is applied along the initial muon-spin direction. Contrary to what is expected for fluctuating magnetism, a clear recovery of the full polarization value for $B_{\parallel} \gtrsim 100$\,mT (see Fig.~\ref{fig:asymmetryLF10}) confirms the static nature of the magnetically ordered phase. The order of
magnitude of the locking field (hundreds of mT) is typical of the static internal fields normally found in pnictides.

\begin{table}[thb]
\centering
\renewcommand{\arraystretch}{1.2}
\caption{\label{tab:TCTN} Magnetic properties of \sfmaof\ samples, as determined from $\mu$SR and dc magnetometry measurements (see text for details).}
\begin{ruledtabular}
\begin{tabular}{p{1mm}lccccc}
\vspace{2mm}
& \lower 0.4mm \hbox{$x$(Mn)}
& \lower 0.4mm \hbox{$T_{\mathrm{N}}$\,(K)}
& \lower 0.4mm \hbox{$\Delta T_{\mathrm{N}}$\,(K)}
& \lower 0.4mm \hbox{$T_{\mathrm{c}}$\,(K)}
& \lower 0.4mm \hbox{$\Delta B_{\mathrm{\mu}}$\,(mT)}\\[2pt]
\hline
&0    &   0       &   --        &  48(1) & --          \\
&0.01 &   0       &   --       &  37(1) & --         \\
&0.03 &  12.5(4)&  3.1(7)  &  21(1) & 24.9(7)  \\
&0.06 &  12.8(2)&  3.3(2)  &  11(1) & 50(2)      \\
&0.08 &  13.5(3)&  3.0(4)  &    --   & 45(5)      \\
&0.10 &  23.3(3)&  12(1)     &    --   & 62(6)      \\
&0.15 &   23(2)   &   15(2)    &    --   & 56(4)      \\
\end{tabular}
\end{ruledtabular}
\end{table}
By considering the $T_{\mathrm{N}}$ and $T_c$ values inferred from the dc-magnetization and \musr\ data, we can draw a tentative phase diagram that describes the evolution of both the SC and M phases as a function of Mn content. As shown in Fig.~\ref{fig:PhaseDiag}, the main feature of the phase diagram is the presence of a narrow region, where both \textit{bulk} superconductivity and FeAs magnetic order coexist over the \textit{whole} sample volume. As established also for other members of the 1111 family (Sm-1111,\cite{Sanna2009} Ce-1111,\cite{Sanna2010,Shiroka2011}, La-1111,\cite{Lamura2014} and Nd-1111\cite{Lamura2015}), the simultaneous presence of M and SC bulk phenomena is compatible with their coexistence at a \emph{nanometer length scale}.
In Fig.~\ref{fig:PhaseDiag} we report also the internal field width $\Delta B_{\mu}$, as determined from the decay
of the transverse component of ZF-$\mu$SR data at 5 K [see Eq.~(\ref{eq:spin_prec})]. $\Delta B_{\mu}$
not only is of the same order of magnitude as the field widths already reported for other 1111 compounds
but, most importantly, its magnitude (proportional to the magnetic order parameter) scales almost regularly
with Mn content. This fact strongly suggests that the magnetic Mn ions induce and
stabilize the magnetically ordered phase.

Finally, regarding the exchange interactions among the Mn ions in the Sm-1111 case we note that: (i) Transport measurements show a residual resistivity that scales with Mn content, with the overall behavior denoting a \emph{decrease} in electronic correlation when compared with the La case.\cite{Sato2010} (ii) The critical temperature $T_c$ vanishes at $x = 0.08$, with a decreasing rate which is in good agreement with the theoretical predictions (squares in Fig.~\ref{fig:PhaseDiag}).\cite{Gastiasoro2016} (iii) 
The magnetically-ordered phase
shows systematically lower $T_{\mathrm{N}}$ values than that of La-1111 compounds, yet of the same order of magnitude as in the La-Y system.\cite{Moroni2016} All these features agree with the theoretical model reported in Ref.~\onlinecite{Gastiasoro2016}, where the rate of suppression of $T_c$ and the type of magnetic order in the Mn-doped 1111 compounds were calculated starting from a multiband superconductor with an $s_{\pm}$ gap symmetry. For the case of magnetic disorder, the $T_c$ suppression rate is not dependent of the assumption of $s_{\pm}$ symmetry, but rather set by the strength of the scatterers and the electronic correlations in the bulk which can enhance the RKKY exchange interactions between Mn ions. In particular, in the Sm-1111 case, the exchange coupling between Mn moments and the conduction electrons is approximately 25\% lower than in the La-1111 system, hence justifying a higher critical impurity concentration of about 8\%.\cite{Gastiasoro2016} The weaker coupling could reflect the smaller ionic size of Sm with respect to La, which implies a smaller unit cell. That is responsible for a lower hopping parameter \textit{t} and therefore for lower electronic correlation effects. Consequently, the Sm-1111 case mimics that of La-Y compounds, where the superconductivity is suppressed at a tenfold value of Mn doping with respect to the pure La-1111 case. Furthermore, a lower magnetic-exchange coupling among the Mn ions implies a lower magnetic ordering temperature (at the same Mn concentration) with respect to the La-1111 system: this is the case of both La-Y-1111 and Sm-1111, whose $T_{\mathrm{N}}$ values are of the same order of magnitude.
\section{\label{sec:concl}Conclusion}
We considered the role of magnetic Mn-for-Fe substitutions in the optimally-doped superconducting compound SmFeAsO$_{0.88}$F$_{0.12}$, a well known member of the 1111 class of iron-based superconductors.
By means of magnetometry and $\mu$SR measurements we could determine the critical superconducting temperature $T_c$ and the magnetic ordering temperature $T_{\mathrm{N}}$, respectively, in samples ranging in Mn content from 0 to 0.15. This allowed us to construct the phase diagram of the \sfmaof\ family and to follow the evolution of the superconductivity from its optimum, achieved at $x=0$, to its extinction at $x=0.08$, and beyond.

Although superconductivity is suppressed only at $x=0.08$, a concomitant AF phase appears already at $x=0.03$, first coexisting at the nanoscale level with SC, then as an increasingly dominant phase, to become the only one above $x=0.08$. While at low Mn substitution rates we observe mostly a depression of the superconductivity, at higher Mn values the cooperative effects among Mn ions reinforce the tendency towards antiferromangetic order. 
The above mentioned findings are fully compatible with a model superconducting system having an $s_{\pm}$ or $s_{++}$ gap-symmetry and moderate electron correlations.\cite{Gastiasoro2016}
\begin{acknowledgments}
This work was performed at the Swiss Muon Source S$\mu$S, Paul Scherrer Institut (PSI, Switzerland) and was in part supported by the Schweizerische Nationalfonds zur F{\"o}rderung der Wissenschaftlichen Forschung (SNF) and the NCCR research pool MaNEP of SNF. G.L.\ and M.P.\ thank M.\ Lucaccini and G.\ Tavilla for their valuable technical support and C.\ Robustelli for the supply of cryogenic liquids. We acknowledge financial support by FP7-EU project SUPER-IRON (No.\ 283204) -M.P.\ and G. L.,  by MIUR project FIRB2012 - HybridNanoDev (Grant No.\ RBFR1236VV) -G.L., and by MIUR project PRIN2012 (Grant No.\ 2012X3YFZ2) -M.P. , G. L., S. S., M. M., P.C., R.D.R.\ and S.B.\\
\end{acknowledgments}

%\bibliography{biblio_Mn}


\end{document}